# Cerebral oxygen extraction fraction MRI: techniques and applications

Dengrong Jiang[1], Hanzhang Lu[1,2,3]

*[1]The Russell H. Morgan Department of Radiology & Radiological Science, Johns Hopkins University School of Medicine, Baltimore, Maryland, USA.*

*[2]Department of Biomedical Engineering, Johns Hopkins University School of Medicine, Baltimore, Maryland, USA.*

*[3]F.M. Kirby Research Center for Functional Brain Imaging, Kennedy Krieger Research Institute, Baltimore, Maryland, USA.*

Corresponding author:
Hanzhang Lu, Ph.D.
The Russell H. Morgan Department of Radiology and Radiological Science
Johns Hopkins University School of Medicine
600 N. Wolfe Street, Park 322
Baltimore, MD 21287
Email: hanzhang.lu@jhu.edu
Phone: 410-955-1431
Fax: 410-614-1977

Grant Sponsors: NIH R01 NS106702, NIH R01 NS106711, NIH R01 AG064792, NIH RF1 AG071515, NIH RF1 NS110041, NIH R21 AG058413, and NIH P41EB031771.

Word count: 7432

# Submitted to Magnetic Resonance in Medicine

## Abstract

The human brain constitutes 2% of the total body mass, but consumes 20% of the oxygen. The rate of the brain's oxygen utilization can be determined from the knowledge of cerebral blood flow and oxygen extraction fraction (OEF). Therefore, OEF is a key physiological parameter of the brain's function and metabolism. OEF has been suggested to be a useful biomarker in a number of brain diseases. With recent advances in MRI techniques, several MRI-based methods have been developed to measure OEF in the human brain. These MRI OEF techniques are based on $T_2$ of blood, phase of blood signal, susceptibility of blood-containing voxel, effect of deoxyhemoglobin on signal behavior in extravascular tissue, and calibration of BOLD signal using gas-inhalation. Compared to $^{15}O$ positron emission tomography, which is considered the "gold standard" for OEF measurement, MRI-based techniques are non-invasive, radiation-free, and have broader availabilities. This article provides a review of these emerging MRI-based OEF techniques. We first briefly introduce the role of OEF in brain oxygen homeostasis. We then review the methodological aspects of different categories of MRI OEF techniques, including their signal mechanisms, acquisition methods, and data analyses. Advantages and limitations of the techniques are discussed. Finally, we review key applications of these techniques in physiological and pathological conditions.

# 1. Introduction

The human brain constitutes 2% of the total body mass, but consumes 20% of the oxygen.[1] The brain has limited capacity to store oxygen, and its oxygen utilization primarily relies on the extraction from incoming arterial blood in real time. Therefore, oxygen extraction fraction (OEF) is a key physiological parameter of the brain's energy metabolism, and has been suggested to be a potential biomarker in several diseases, such as Alzheimer's disease (AD),[2,3] carotid steno-occlusive disease,[4] sickle cell disease (SCD)[5,6] and brain tumor.[7]

Measurement of OEF and the related cerebral metabolic rate of oxygen ($CMRO_2$) in humans used to be a niche market of positron emission tomography (PET) with $^{15}O$-labeled radiotracers.[8-10] Although $^{15}O$-PET is still widely regarded as the gold standard for OEF and $CMRO_2$ mapping, its broad clinical applications have been hampered by the complex logistics, exposure to radiation, and the requirement for an onsite cyclotron to produce the $^{15}O$ isotope, which has a short half-life of 2 minutes.

With the advances in MRI, several techniques have been developed to quantify OEF, based on the associations between blood oxygenation and MRI properties such as $T_2$,[11-13] susceptibility[14-17] and magnetization phase,[18-21] or by exploiting gas modulations.[22-24] Among these techniques, some provide a global or whole-brain measure of OEF, while others aim to estimate OEF in specific brain regions. Some of these techniques have demonstrated potential clinical utility in brain diseases. The goal of this article is to provide a review of MRI-based OEF measurements. We will first introduce the physiological importance of OEF. We will then review major categories of MRI-based OEF techniques, including their signal mechanisms, acquisition methods, and data analyses. Finally, we will highlight key clinical applications of these techniques, although they are not meant to be exhaustive.

# 2. OEF and brain oxygen homeostasis

As illustrated in Figure 1, when fully oxygenated arterial blood flows through the capillaries, it releases oxygen to the surrounding tissues. The fraction of oxygen extracted by the tissue is OEF. Specifically, OEF is defined as:

$$OEF = \frac{Y_a - Y_v}{Y_a} \times 100\% \qquad [1]$$

where $Y_v$ is the venous oxygenation, defined as the fraction of oxyhemoglobin in the venous blood, and $Y_a$ is the arterial oxygenation. Under normal conditions, $Y_a \approx 100\%$ and is relatively uniform throughout the body. Thus, OEF can often be simplified to:

$$OEF \approx 1 - Y_v \qquad [2]$$

$Y_a$ can also be conveniently measured by a pulse oximeter at the fingertip, especially when abnormalities in $Y_a$ is expected. The major challenge is therefore the measurement of $Y_v$.

In blood, oxyhemoglobin and deoxyhemoglobin have distinct magnetic properties: oxyhemoglobin is diamagnetic, while deoxyhemoglobin is paramagnetic. Inside the vessel, increased concentration of deoxyhemoglobin causes a reduction in blood $T_2$ through water diffusion and exchange, similar to the well-known BOLD effect.[11-13] Higher concentration of deoxyhemoglobin also increases the magnetic susceptibility of voxels containing blood.[14,25] This susceptibility difference between blood and tissue results in an additional phase angle in the transverse magnetization of blood.[18-21] Outside the blood vessel, the local field inhomogeneity induced by the paramagnetic deoxyhemoglobin in the vessel networks results in additional decay of the tissue signal.[26,27] Finally, BOLD signal changes during hypercapnic and hyperoxic gas inhalations can be modeled as a function of deoxyhemoglobin, and the non-linear nature of this function allows one to tease apart different variables and estimate OEF (among other physiological parameters).[22-24] The OEF methods described in this review are based on one or more of the above-mentioned effects.

Once OEF is quantified, it can be combined with CBF to calculate $CMRO_2$ based on the Fick's principle:

$$CMRO_2 = CBF \cdot (Y_a - Y_v) \cdot C_{hb} \cdot [Hb] = CBF \cdot OEF \cdot Y_a \cdot C_{hb} \cdot [Hb] \qquad [3]$$

where [Hb] is the total hemoglobin concentration (in gram hemoglobin/dL blood). $C_{hb}$ is the oxygen carrying capacity of hemoglobin (1.34mL or 59.8μmol $O_2$/gram hemoglobin).[28] The term $(Y_a - Y_v) \cdot C_{hb} \cdot [Hb]$ is sometimes called the arterio-venous difference in oxygen content ($AVDO_2$). CBF can be measured globally by using phase-contrast MRI to quantify blood flow in the feeding arteries[29] or regionally by using arterial spin labeling (ASL) MRI.[30]

It is worth noting that non-invasive quantification of OEF and $CMRO_2$ is also possible with MR spectroscopic imaging of $^{17}O$, which is the only MR active stable isotope of oxygen. Due to the very low natural abundance of $^{17}O$, $^{17}O$ MRI is typically performed with $^{17}O$-enriched tracers.[31-36] Detailed review of $^{17}O$ MRI is beyond the scope of this article, and interested readers are referred to other dedicated reviews.[34-36] In this article, we will focus on MRI OEF techniques that do not require exogenous tracers.

# 3. MRI techniques for OEF measurement

## 3.1. $T_2$-based methods

### *3.1.1. Measurement of venous $T_2$*

Under a fixed hematocrit (Hct) level, the blood oxygenation has a one-to-one correspondence to blood $T_2$.[11-13] Therefore, $Y_v$ (and OEF) can be measured by quantifying its $T_2$.[37,38] It is important to obtain pure venous blood signal in this measurement, as partial volume contamination from tissue or CSF can cause bias in $T_2$ determination. While some rely on high image resolution to separate blood signal in large cerebral veins,[39] more recent work sought to use special sequence design to isolate blood signal.

One example is $T_2$-relaxation-under-spin-tagging (TRUST) MRI.[40-42] As shown in Figure 2, TRUST labels blood spins on the venous side (Figure 2A), and uses the subtraction between control and labeled images (Figure 2B) to yield pure venous blood signal in the superior sagittal sinus (SSS). The venous blood $T_2$ is quantified by applying $T_2$-preparation with varying number of refocusing pulses, which is characterized by the effective TE (eTE).[40] The $T_2$-preparation pulses were applied to the entire brain, i.e., non-selectively, so that the degree of $T_2$ attenuation is not dependent on the flow velocity of the blood. Simple mono-exponential fitting of the blood signal as a function of eTE then yields venous $T_2$, which is converted to $Y_v$ through a calibration model, based on the Hct level of the subject.[43] Since SSS drains the majority of cerebral cortex, TRUST MRI is thought to assess the global OEF level of the brain.

TRUST has a scan time of 1.2 min, and its signal mechanism is relatively straightforward. Rater dependence in TRUST data processing is minimal, as it uses the top signal voxels for $T_2$ estimation and the manual region-of-interest (ROI) plays a negligible role in the final result as long as it contains the SSS.[40] Test-retest reproducibility of TRUST MRI has been examined, with a same-day test-retest coefficient-of-variation (CoV) of 3%[44,45] and a day-to-day CoV of 8% for OEF quantification.[44] The sensitivity of TRUST to OEF changes has been evaluated in several physiological challenges, such as caffeine ingestion,[46] hypercapnia,[47] hypoxia and hyperoxia,[48] all of which showed expected OEF changes using TRUST. The reliability of TRUST MRI has also been studied in the context of multiple sites[49] and different MR vendors.[45] In a recent study,

TRUST MRI was validated with $^{15}O$-PET, which found a global OEF of 36.44±4.07% with TRUST MRI and 36.45±3.65% with $^{15}O$-PET. The two modalities also revealed a correlation (intraclass correlation coefficient = 0.90) in OEF measures.[41] Besides human studies, TRUST has also been implemented on animal MR imaging systems for OEF assessment in mice and rats.[50-53]

Other interesting technical work on global $T_2$ assessment include the use of high-spatial-resolution echo-planar or turbo-field-echo acquisition to obtain pure blood voxels,[54,55] the combination of $T_2$-preparation sequence with inversion recovery to simultaneously measure blood $T_1$,[56,57] and the integration of flow measurement in the sequence to allow $CMRO_2$ measurement from a single sequence.[58] While global OEF estimation has the obvious limitation of lack of spatial information, these techniques tend to have a high reproducibility with a short scan duration, which makes it easy for them to be incorporated into the workflow of clinical or research imaging. The global OEF can also serve as a useful reference or validation measure for comparison with regional OEF techniques.

$T_2$-based regional OEF techniques have also been reported. O'Brien et al. proposed to incorporate spatially selective saturation pulses into TRUST to isolate venous blood signal from localized regions of the brain, so that OEF can be measured for a specific hemisphere or a volume of interest.[59] Another technique, $T_2$-relaxation-under-phase-contrast (TRUPC), employs phase-contrast complex subtraction to isolate flowing blood signal in the vessels.[60,61] With this technique, OEF can be measured in regional cerebral veins, such as internal cerebral vein (ICV), the great vein of Galen, straight sinus, as well as pial veins.[60-62] TRUPC has a same-day CoV of 6% for OEF quantification.[60] An accelerated version of TRUPC has also been proposed,[62] which shortens the acquisition time from 5 minutes to approximately 1 minute. TRUPC MRI has been validated against blood-gas oximetry of direct blood samples from the SSS on a piglet model.[63] Three-dimensional and $T_2^*$-based implementations of TRUPC have also been demonstrated.[64]

Further advancement of $T_2$-based methods attempted to provide voxel-wise OEF mapping. QUantitative Imaging of eXtraction of Oxygen and TIssue Consumption (QUIXOTIC)[65] and Velocity-Selective Excitation with Arterial Nulling (VSEAN)[66] are two techniques that use velocity-selective labeling or excitation to isolate post-capillary venous blood signal from tissues in the same voxel, allowing voxel-wise mapping of venous blood $T_2$ and OEF. However, since the post-capillary venous blood only comprises 1-3% of the volume of a voxel,[67] signal-to-noise-ratio (SNR) and partial volume contamination represent potential obstacles in clinical applications.

Approaches to shorten scan time,[68] enhance SNR and reduce CSF contamination[69] have been proposed.

### 3.1.2. $T_2$-Y conversion

All $T_2$-based oximetry methods require the use of a calibration model to convert the MR property of blood $T_2$ to the physiological parameter of oxygenation (Y). It should be noted that, under most circumstances, the same calibration plot can be used for different subjects and different studies. That is, one does not need to develop a calibration plot for each subject or each study. For applications in children and adults that have no hematological pathologies, the most commonly used $T_2$-Y calibration model was proposed by Lu et al. (Figure 3),[43] which was based on in vitro experiments on bovine blood samples. Bovine blood has similar characteristics to adult human blood in terms of the hemoglobin structure, red blood cell (RBC) shape and size,[70] and diffusional water permeability.[71] This bovine model has been demonstrated to provide $Y_v$ and OEF quantifications consistent with $^{15}$O-PET on healthy human adults.[41]

Some studies also considered the effect of Hct on the $T_2$-Y relationship. Blood $T_2$ is indeed also dependent on Hct, although to a much less degree compared to oxygenation.[43] To take Hct into consideration in the conversion, one can use the calibration model to obtain Hct-specific $T_2$-Y curve (most models provide a three-way relationship of Y, Hct, and $T_2$; thus, it is straightforward to obtain Hct-specific curves). Then, based on the individual's Hct value which is sometimes available in clinical workup or can be measured from a separate blood sampling or MRI sequence,[72,73] one can use the proper $T_2$-Y curve for the conversion (Figure 3B). When the individual Hct of the subject is unavailable, some studies have used 0.42 for males and 0.40 for females,[74,75] accounting for some of the sex-related differences in Hct.

One limitation of the bovine model is that it was calibrated with a limited Hct range of 35%-55%.[43] Although this range is sufficient for most healthy volunteers and patients, extrapolation to anemic patients with very low Hct is somewhat controversial.[63,76] This model may also be inappropriate for populations with atypical hemoglobin or RBCs. For example, in human neonates, the major type of hemoglobin is fetal hemoglobin (HbF), which has different structures than the adult hemoglobin (HbA).[77] A neonate-specific $T_2$-Y calibration model has been reported.[78] Another example is SCD. Patients with SCD have sickle hemoglobin (HbS), which is highly unstable especially in its deoxygenated form, causing hemoglobin polymerization and precipitation as well as RBC deformation (sickle appearance).[79] SCD patients can also have a considerable amount of HbF.[54,80] Therefore, the magnetic property of sickled blood may be different from normal blood. SCD-specific $T_2$-Y calibration models have been reported.[5,54,81] There is still much controversy in this area, but some of the SCD models have shown a substantial difference from the bovine model.[5] One of the models further takes into account the varying percentage of HbS among SCD patients, especially those undergoing transfusion therapies.[81] However, it should be pointed out that, while the differences associated with hemoglobin types have been hypothesized to be the main reason underlying disparities across models, it is also possible that experimental preparations (e.g. how temperature, pH, and cell integrity of the in vitro blood are controlled) and pulse sequences used (e.g. turbo-field-echo $T_2$ suffers from variations in magnetization across k lines and may under-estimate $T_2$ in SCD[54]) may have contributed to some of these differences.

Future work should consider individual-specific, in vivo calibrations for patients with hematological conditions such as SCD. Such calibration can be conducted by applying MRI $T_2$ measurements to the superficial veins in the arm, for example, followed by blood sampling to obtain its oxygenation. A two-point (or multi-point) calibration can be further achieved by inducing oxygenation changes using arm compression to restrict blood flow or hyperoxic/hypoxic challenge. Such calibration data can then be applied to the brain venous $T_2$ to provide a more accurate estimation of OEF in the patient. Compared to previous calibration efforts, this approach can preserve erythrocytes' cell integrity and their nature shape in vivo, and will not have confounding factors related to blood preparation.

Finally, the above-mentioned models were primarily calibrated at 3T, with some conducted at 7T.[82] Since $T_2$ varies with magnetic field strength, for applications at other field strengths, specific calibration models should be used.[13,82-84] A unified calibration model based on quantitative theory and aggregated experiments has also been proposed.[13]

### 3.2. Phase-based methods

Because of the paramagnetic property of deoxyhemoglobin, the susceptibility difference ($\Delta\chi$) between blood and tissue has a linear relationship with the blood oxygenation:[14,85,86]

$$\frac{\Delta\chi}{Hct} = \Delta\chi_{do}(1 - Y) + \Delta\chi_{oxy} \quad [4]$$

where $\Delta\chi_{do}$ is the susceptibility difference between fully oxygenated and deoxygenated blood, and has been characterized in the literature.[14,80,85-87] $\Delta\chi_{oxy}$ is the susceptibility difference between fully oxygenated blood and tissue. $\Delta\chi_{oxy}$ is much smaller than $\Delta\chi_{do}$[85] and is often omitted in many studies. The susceptibility of blood induces local magnetic field offset $\Delta B_{bl}$, which results in an additional phase difference $\Delta\Phi$ between successive echoes. Although in general the inverse problem to solve $\Delta\chi$ from $\Delta\Phi$ is complex and ill-conditioned,[88,89] in a special case where the blood vessel can be modeled as a long straight cylinder approximately parallel to the main magnetic field $B_0$, we have:[25]

$$\Delta\Phi_{ie} = \Delta\varphi_{intra} - \Delta\varphi_{extra} \qquad [5]$$

where $\Delta\varphi_{intra}$ and $\Delta\varphi_{extra}$ are the inter-echo phase difference within the vessel and in the neighboring extravascular tissue, respectively, and can be written as:

$$\Delta\varphi_{intra} = \gamma \cdot \Delta TE \cdot \left(\Delta B_{bl,intra} + \Delta B_{0,intra}\right) \qquad [6]$$

$$\Delta\varphi_{extra} = \gamma \cdot \Delta TE \cdot \left(\Delta B_{bl,extra} + \Delta B_{0,extra}\right) \qquad [7]$$

where $\gamma$ is the gyromagnetic ratio and $\Delta TE$ is the TE difference between successive echoes. $\Delta B_{0,intra}$ and $\Delta B_{0,extra}$ are the intra- and extravascular background $B_0$ inhomogeneity, respectively. $\Delta B_{bl,intra}$ and $\Delta B_{bl,extra}$ are the blood-induced field inside and outside the vessel, respectively, and can be written as:[25]

$$\Delta B_{bl,intra} = \frac{\Delta\chi}{2}\left(cos^2\theta - \frac{1}{3}\right)B_0 \qquad [8]$$

$$\Delta B_{bl,extra} = \frac{\Delta\chi}{2}\left(sin^2\theta \cdot \frac{r^2}{\rho^2} \cdot cos2\omega\right)B_0 \qquad [9]$$

where $\theta$ is the angle between the cylinder axis and $B_0$, $r$ is the vessel radius, $\rho$ is the distance to the vessel axis, and $\omega$ is the azimuthal angle (Figure 4a). Assuming that the intra- and extravascular ROIs are sufficiently adjacent to each other thus the effects of background $B_0$ inhomogeneity, i.e., $\Delta B_{0,intra}$ and $\Delta B_{0,extra}$, are the same, $\Delta\Phi_{ie}$ is primarily related to the deoxyhemoglobin content in the blood. Figure 4b shows a simulated blood-induced field map (Eqs. [8,9]) for a hypothetical cylindric vessel oriented at $\theta$=20.1°.[90] It can be seen that the extravascular term, $\Delta B_{bl,extra}$, is very small when $\theta$ is small, and also decays rapidly with the distance from the vessel. Therefore, in most experimental studies where the vessel is approximately parallel to $B_0$, $\Delta B_{bl,extra}$ is negligible and Eq. [5] can be written as:

$$\Delta\Phi_{ie} = \frac{\Delta\chi}{2}\left(cos^2\theta - \frac{1}{3}\right)B_0\gamma\Delta TE \qquad [10]$$

Note that θ can be measured from localizer or time-of-flight (TOF) vessel images.[91,92] Combining Eqs. [4,10] and ignoring $\Delta\chi_{oxy}$, $Y_v$ can be calculated by:[25]

$$Y_v = 1 - \frac{2|\Delta\Phi_{ie}|}{\gamma B_0 \cdot \Delta TE \cdot \Delta\chi_{do} \cdot Hct\left(cos^2\theta - \frac{1}{3}\right)} \qquad [11]$$

To quantify OEF using the phase-based method, gradient-recalled-echo (GRE) sequences with multiple TEs are typically used.[18,19] Flow compensation is usually applied to minimize the flow-induced phase accumulation. Because the vessel is assumed to be a long straight cylinder, phase-based methods usually focus on large cerebral veins that are relatively parallel to $B_0$, such as internal jugular veins (IJVs) and SSS,[19,20,93] providing an estimation of global $Y_v$ (and OEF) similar to the TRUST method. Figure 5 shows exemplar data of phase-based OEF at the level of SSS.[94] Some studies have also extended this technique to quantify OEF in selected regional veins that can be approximated as a long cylinder (length to diameter ratio > 5).[18,21,95,96]

The accuracy of phase-based $Y_v$ quantification relies on the calibration of $\Delta\chi_{do}$. For normal human adult blood, early works found a $\Delta\chi_{do}$ around 0.18 ppm (cgs units),[14] while later studies reported $\Delta\chi_{do}$ to be 0.27 ppm (cgs).[85,86] A recent study by Eldeniz et al. reported that $\Delta\chi_{do}$ was 0.19 ppm (cgs) for normal adult blood and the $\Delta\chi_{do}$ of sickled blood was not significantly different.[80] For human neonatal blood, Portnoy et al. found a $\Delta\chi_{do}$ of 0.21 ppm (cgs).[87]

The sensitivity of phase-based methods to OEF changes have been demonstrated in various physiological challenges, such as hypercapnia,[93] hypoxia,[97] acetazolamide administration,[98,99] caffeine ingestion[99] and breath-hold apnea.[92,100]

Phase-based methods have been integrated with phase-contrast MRI to simultaneously measure global OEF and CBF levels in a single sequence dubbed OxFlow.[20] OxFlow has a CoV of 5% for global OEF quantification in same-day and day-to-day test-retest experiments.[20,94] Recently, OxFlow has been demonstrated to provide global OEF values with no systematic bias compared to $^{15}$O-PET measurements on a porcine model.[101] The original OxFlow sequence took 30 seconds.[20] With extensive acceleration, the variants of OxFlow can achieve a temporal resolution of a few seconds.[92,100,102]

It should be noted that actual cerebral veins often exhibit curvature and noncircular cross-sections, thus rarely conform to the straight cylinder model. For example, SSS has a triangular cross-section. Fortunately, it has been demonstrated that the absolute error in $Y_v$ caused by non-circular cross-section is < 5% when the vessel tilt angle is small ($\theta$ < 30°).[20,90,91] Alternatively, Driver et al. proposed to use a forward field calculation that takes into account the exact shape of the vessel to estimate $Y_v$.[103] Another major source of error is the phase errors induced by background $B_0$ inhomogeneities, which can be approximated as a second-order polynomial and then removed from the phase difference calculations in Eqs. [6,7].[104] In addition, to minimize partial-volume effects, sufficient image resolution is required. Finally, the processing of the data requires the manual drawing of extra- and intravascular ROIs, which need to be approached carefully to minimize rater dependence.

### 3.3. Methods based on quantitative susceptibility mapping

To estimate OEF in local veins with arbitrary orientation and geometry, quantitative susceptibility mapping (QSM) methods have been developed. QSM solves the general inverse problem to determine regional $\Delta\chi$ from $\Delta\Phi$. Under the assumption that the vein is at least one-voxel wide thus the $\Delta\chi$ measured is linearly proportional to blood oxygenation but not due to blood volume fraction, regional $Y_v$ can be estimated using Eq. [4]. As mentioned earlier, this inverse problem is complex and ill-conditioned.[88,89] Various algorithms have been proposed to solve this problem and quantify $Y_v$ and OEF in local veins, such as thresholded *k*-space division[15] and L1-regularization.[16,105] To further correct for the flow-induced phase errors in the veins, an adaptive-quadratic fitting method has been proposed.[106] A potential caveat of these methods is that they may be sensitive to partial volume effects, i.e., the voxel is not 100% blood, which will result in an overestimation in venous oxygenation (thus underestimation in OEF), especially for small veins.[15,16,25,107] Partial volume correction algorithms may be useful in improving the accuracy of OEF quantification.[108,109]

Besides local veins, QSM-based methods have been extended to measure voxel-wise OEF.[17,110] This method does not require the vein(s) to occupy the entire voxel. For an arbitrary voxel in the brain, its susceptibility has contributions from various sources including blood and non-blood tissues. To decouple these factors, one can acquire QSM and CBF (e.g., using ASL) in two isometabolic brain states to correct for the non-blood contributions (e.g., from tissue ferritin), which are assumed to remain the same between the two states. This has been carried out by using physiological challenges such as caffeine ingestion,[17] hyperventilation,[110] hypercapnia[111] or hyperoxia.[112] These challenges are generally considered to be isometabolic (i.e., does not change $CMRO_2$), although controversial findings exist in the literature.[46-48,58,93,113-117] The blood contribution arises from both venous and arterial blood and is modulated by their respective cerebral blood volume (CBV).[17,110] In these methods,[17,110-112] total CBV was estimated from CBF using relationships reported in the PET literature,[118,119] and a certain ratio between arterial CBV and venous CBV ($CBV_v$) was assumed to separate out the venous susceptibility. Then, OEF at baseline and during physiological challenges can be quantified.[17,110-112] Further development of QSM-based methods attempted to map OEF without any physiological challenges. For example, Zhang et al. proposed to estimate OEF from baseline QSM and CBF data by assuming $CMRO_2$ to have minimum local variance.[120] A caveat of these methods is the large number of assumptions, e.g., CBF-CBV relationship, arterial-venous CBV fractions, isometabolic challenge, and registration between EPI-based low-resolution ASL and GRE-based high-resolution QSM images. Additionally, the accuracy of the measurement is expected to strongly dependent on the reliability of the ASL CBF estimation in both baseline and challenged states, in the presence of confounding factors such as labeling efficiency, bolus arrival time, and arterial contaminations.[121-124]

A few studies have compared QSM-based OEF measurements to PET-based OEF in healthy volunteers and in patients with arterial steno-occlusive diseases,[125-128] in which the affected hemisphere is likely to have an elevated OEF. These studies showed significant correlations between QSM and PET in both the absolute OEF values and the affected/contralateral OEF ratio, although systematic biases were also reported.[125-127]

### 3.4. Quantitative BOLD

Unlike the $T_2$-based and susceptibility-based methods that focus on the effect of blood oxygenation on “intravascular” signals, the quantitative BOLD (qBOLD) model focuses on the signal decay in the “extravascular” space due to the local field inhomogeneities induced by the paramagnetic deoxyhemoglobin in the vessel network. When studying the transverse signal decay due to deoxyhemoglobin, conventional BOLD models consider that the decay rate is related to both oxygenation level (i.e., OEF) and blood volume, but are unable to separate the effects of these two terms. That is, a faster BOLD signal decay could be due to either lower blood oxygenation or larger blood volume. Yablonskiy and colleagues, on the other hand, proposed a “static dephasing regime” BOLD model, in which the vessels were considered as an ensemble of randomly oriented cylinders.[26] The key advantage of this model is that the oxygenation and blood volume effects can be separated experimentally. Specifically, two asymptotic equations can be derived to describe the signal behavior around a spin-echo:[26,129,130]

$$\ln\left(S_S(\tau)\right) = \ln\left(S(0)\right) - (TE+\tau)\cdot R_2 - 0.3\cdot DBV\cdot(\delta\omega\cdot\tau)^2, \text{ if } |\tau| < 1.5/\delta\omega \quad [12]$$

$$\ln\left(S_L(\tau)\right) = \ln\left(S(0)\right) - (TE+\tau)\cdot R_2 - R_2'|\tau| + DBV, \text{ if } |\tau| > 1.5/\delta\omega \quad [13]$$

where τ is the time gap between image acquisition and the spin-echo (τ is negative if the image is sampled before the spin-echo); $S_S$ and $S_L$ are the extravascular tissue signal for short and long τ, respectively; $R_2$ is the tissue transverse relaxation rate; DBV is the deoxygenated blood volume, which is mainly contributed to by the veins but also contains the part of capillary adjacent to the venous side. TE here is the time of the spin-echo. δω is the deoxyhemoglobin induced frequency shift; $R_2'$ is the reversible transverse relaxation rate ($R_2' = R_2^* - R_2$) and has the following relationship:[26,129,130]

$$R_2' = DBV\cdot\delta\omega = DBV\cdot\gamma\cdot\frac{4}{3}\pi\cdot\Delta\chi_{do}\cdot Hct\cdot(1-Y_v)\cdot B_0 \quad [14]$$

Figure 6 shows a schematic of the qBOLD model.[130] The linear part of the model (Eq. [13]) is used to measure $R_2'$, while the mismatch between the fitted linear intercept and the spin-echo signal allows estimation of DBV. OEF can then be calculated using Eq. [14].

To quantify OEF using the qBOLD model, many studies used a Gradient-Echo-Sampling-of-Spin-Echo (GESSE) sequence, in which multiple gradient echoes are acquired with varying τ from the spin-echo of a fixed TE.[129,131-133] Figure 7 shows exemplar $R_2'$, DBV and OEF maps acquired with GESSE-based qBOLD.[133] However, one drawback of this sequence is the varying $R_2$-weighting among the gradient echo signals (Eq. [13]), which must be accounted for when quantifying $R_2'$.[131-133] Alternatively, An et al. proposed an Asymmetric-Spin-Echo (ASE) sequence in which the position of the 180° refocusing pulse is varied to yield different τ while keeping (TE+τ) constant. This allows direct estimation of $R_2'$ without the confounding $R_2$-weighting effects.[134]

Both GESSE and ASE sequences allow OEF mapping within approximately 10 minutes.[130-134] Reduction of scan time by using fast readout schemes has been proposed.[135] Sensitivity of ASE to OEF changes has been shown in carbogen inhalation challenge.[136] A recent study evaluated the test-rest reproducibility of an ASE-based method and showed a same-day CoV of < 2% and day-to-day CoV of <3% for OEF averaged across gray matter, white matter, or whole-brain.[137] A few studies have evaluated the accuracy of qBOLD-based $Y_v$ and OEF measurements on rodent models, by comparing the $Y_v$ in the SSS estimated by blood gas oximetry with the whole-brain averaged qBOLD-based $Y_v$.[138-140] These studies demonstrated correlations between blood gas oximetry and qBOLD,[138-140] but some also showed a bias.[140]

Because the separation of the DBV and OEF effects in the qBOLD model relies on subtle differences in decay patterns (Eqs. [12,13]), accurate estimation of DBV and OEF from the GESSE or ASE signals requires high SNR.[129,141,142] To address this problem, several groups developed a multiparametric-qBOLD (mqBOLD) scheme, in which CBV is measured using dynamic susceptibility contrast (DSC) MRI and $R_2'$ is estimated by separately mapping $R_2$ with a multiple spin-echo sequence and $R_2^*$ by a multi-echo GRE sequence.[143-145] However, the CBV measured by DSC is the total CBV rather than DBV. Therefore, the OEF produced by mqBOLD is a relative OEF.[144,145] Alternatively, Stone et al. proposed to measure DBV through a hyperoxia challenge, while $R_2'$ was quantified using ASE.[146,147] In this study, OEF values obtained with separate DBV measurement had substantially better agreement with TRUST global OEF than those obtained using DBV values estimated from the ASE signal.[146,147] This manifested the importance of accurate DBV measurement in qBOLD-based OEF quantification.

A caveat of the qBOLD method is that the model described in Eqs. [12-14] considered a single extravascular tissue compartment and assumed that $R_2'$ is only related to the deoxygenated blood.[26,129] However, in reality, $R_2'$ is also sensitive to macroscopic $B_0$ inhomogeneity, which must be corrected either prospectively using the z-shimming method,[130,148] or retrospectively using a high-resolution field map[133] or by modeling the voxel spread function.[149,150] An actual imaging voxel can also contain other compartments, such as intravascular blood and CSF or interstitial fluid (ISF). To address this issue, He et al. extended the original qBOLD model to incorporate the contributions from other compartments.[133] However, solving this multi-compartmental model requires prior knowledge about the tissue composition and results in a large number of fitting parameters.[133] Alternatively, one can minimize the contributions from other compartments to simplify the model. For example, the intravascular contribution can be reduced by using flow crushing gradients,[134] and the CSF/ISF signal can be suppressed by using a fluid-attenuated-inversion-recovery preparation pulse.[130] Another important assumption of the qBOLD model is "static dephasing", i.e., neglecting the diffusion effects. However, it has been shown that diffusion introduces a vessel size dependent effect on the signal decay, and ignoring this diffusion effect may lead to systematic underestimation of OEF.[151-154]

### 3.5. Dual-calibrated fMRI

BOLD-based fMRI is widely used to investigate the brain's function. It has also been applied to determine OEF when combined with gas inhalation physiological challenges. The $R_2^*$-weighted BOLD signal is dependent on CBV and the concentration of deoxyhemoglobin. Specifically, the relative BOLD signal change can be written as:[155,156]

$$\frac{\Delta BOLD}{BOLD_0} = M\left(1 - \left(\frac{CBV_1}{CBV_0}\right)\left(\frac{[dHb]_1}{[dHb]_0}\right)^{\beta}\right) \quad [15]$$

where the parameter $M$ is a composite constant related to field strength and TE, among other factors, and represents the maximum possible BOLD signal change that can be observed from the voxel. [dHb] is the deoxyhemoglobin concentration. The subscript "0" denotes the baseline values while the subscript "1" denotes challenged state values. The constant β indicates the relationship between deoxyhemoglobin concentration and $R_2^*$, and is dependent on vascular morphology, water diffusion and the field strength, typically assumed to be 1.3 or 1.5 for 3T applications.[22-24,155,157-159] Note that [dHb] has the following relationship with $Y_v$:

$$[dHb] = [Hb](1 - Y_v) \quad [16]$$

where [Hb] is the total hemoglobin concentration.

Eq. [15] is a general expression and is applicable to a variety of challenges such as task activation and physiological maneuvers. The dual-calibrated fMRI (dc-fMRI) OEF method applies this relationship to two gas inhalation regimes, hypercapnia and hyperoxia, to estimate $Y_v$ (thereby OEF).

First, hypercapnia via $CO_2$ inhalation is employed to determine $M$. The MRI data acquisition involves the measurement of both BOLD and CBF (via ASL) at the baseline and during the hypercapnia challenge. The relative CBV change in Eq. [15] is then inferred from the relative CBF change by a power-law model:[119]

$$\frac{CBV_1}{CBV_0} = \left(\frac{CBF_1}{CBF_0}\right)^{\alpha} \quad [17]$$

The constant α was initially measured to be 0.38 for total CBV by Grubb et al.[119] However, it is well known that BOLD signal is primarily modulated by $CBV_v$ rather than total CBV. Later studies have confirmed the power-law relationship for $CBV_v$, but reported a substantially lower α value of ~0.2.[160,161]

Based on Eq. [3] mentioned earlier, assuming $Y_a$=100%, and since [Hb] and $C_{hb}$ are unchanged with hypercapnia, $\frac{[dHb]_1}{[dHb]_0}$ can be rewritten as:

$$\frac{[dHb]_{hc}}{[dHb]_0} = \frac{CMRO_{2,hc}}{CMRO_{2,0}} \cdot \frac{CBF_0}{CBF_{hc}} \qquad [18]$$

where the subscript "hc" denotes the values during hypercapnia. By combining Eqs. [15,17,18], it can be derived that:

$$\frac{\Delta BOLD_{hc}}{BOLD_0} = M\left(1 - \left(\frac{CBF_{hc}}{CBF_0}\right)^{\alpha-\beta}\left(\frac{CMRO_{2,hc}}{CMRO_{2,0}}\right)^{\beta}\right) \qquad [19]$$

Under the assumption that hypercapnia is isometabolic ($CMRO_{2,hc}$=$CMRO_{2,0}$), $M$ can then be estimated based on the hypercapnia-induced BOLD signal change and CBF change:

$$M = \frac{\Delta BOLD_{hc}}{BOLD_0} \cdot \frac{1}{1 - \left(\frac{CBF_{hc}}{CBF_0}\right)^{\alpha-\beta}} \qquad [20]$$

Next, hyperoxia challenge is applied while BOLD and CBF changes are measured. To apply Eq. [15] to hyperoxia, $CMRO_2$ is again assumed to be unchanged and the $\frac{[dHb]_1}{[dHb]_0}$ term is rewritten as:[24]

$$\frac{[dHb]_{ho}}{[dHb]_0} = \frac{CBF_0}{CBF_{ho}} - \frac{1}{[dHb]_0}\left\{\frac{1}{C_{hb}}\left(C_aO_{2,ho} - \left(\frac{CBF_0}{CBF_{ho}}\right)C_aO_{2,0}\right) + [Hb]\left(\frac{CBF_0}{CBF_{ho}} - 1\right)\right\} \qquad [21]$$

where the subscript "ho" denotes the values during hyperoxia. $C_{hb}$ is the oxygen carrying capacity of hemoglobin (59.8μmol $O_2$/gram hemoglobin).[28] $C_aO_2$ is the arterial oxygen content. Here $C_aO_2$ must consider the dissolved oxygen (which is no longer negligible), in addition to hemoglobin-bound oxygen, and can be written as:[24]

$$C_aO_2 = C_{hb} \cdot [Hb] \cdot \frac{1}{\left(\frac{23400}{(P_aO_2)^3 + 150 \cdot P_aO_2} + 1\right)} + P_aO_2 \cdot \varepsilon \qquad [22]$$

where the first term represents $O_2$ bound to hemoglobin while the second term is the dissolved $O_2$ in blood plasma. $P_aO_2$ is the arterial oxygen partial pressure and can be approximated by the end-tidal oxygen partial pressure ($P_{ET}O_2$) measurement. ε is the coefficient of solubility of oxygen in blood (0.138μmol $O_2$/dL blood/mmHg $O_2$ tension).[48,162] Finally, by combining Eqs. [2,15-17,20-22], we have:

$$OEF_0 = \frac{[dHb]_0}{[Hb]} = \frac{\frac{1}{C_{hb}} \cdot \frac{1}{[Hb]} \left( C_a O_{2,ho} - \left( \frac{CBF_0}{CBF_{ho}} \right) C_a O_{2,0} \right) + \left( \frac{CBF_0}{CBF_{ho}} - 1 \right)}{\frac{CBF_0}{CBF_{ho}} - \left( 1 - \frac{\Delta BOLD_{ho}}{BOLD_0 \cdot M} \right)^{\frac{1}{\beta}} \left( \frac{CBF_0}{CBF_{ho}} \right)^{\frac{\alpha}{\beta}}} \qquad [23]$$

Therefore, baseline OEF can be measured by acquiring BOLD signal, CBF and $P_{ET}O_2$ (to estimate $C_aO_2$ per Eq. [22]) at baseline and during two gas challenges: hypercapnia to calibrate *M* and hyperoxia to factor out $[dHb]_0$. A schematic of dc-fMRI experiments is shown in Figure 8.[163] The total duration of a dc-fMRI experiment is typically about 18 min.[22-24,163]

To simultaneously acquire both BOLD and CBF images at the baseline and during gas challenges, one approach is to use an ASL sequence with a relatively long TE (~20ms) to increase the BOLD weighting. BOLD signal is then extracted by averaging adjacent ASL label and control images, while CBF is calculated from the label-control difference.[23] However, the TE used in this approach is suboptimal for both ASL and BOLD, leading to a compromised contrast-to-noise-ratio.[163] Later studies of dc-fMRI have mainly used dual- or multi-echo ASL sequences in which the CBF images are acquired at an early echo with shortest TE to maximize SNR, while the BOLD images are acquired at a later echo with optimal BOLD contrast.[22,24,117,158,164,165] The dual-echo sequence contains greater cross-talks between the CBF and BOLD effects. Thus, some studies have used dual-excitation schemes in which a conventional ASL readout is immediately followed by another excitation to acquire BOLD images.[166-168]

Data processing of dc-fMRI is relatively complicated. One approach is to compute *M* and $OEF_0$ separately using the hypercapnia and hyperoxia data,[23] as described above, but may suffer from error propagations along the analysis pipeline. Alternatively, one can jointly estimate these parameters by fitting a generalized model using all data, which has been suggested to improve the robustness of OEF estimation.[169] Machine-learning based method has also been proposed.[170]

dc-fMRI allows simultaneous measurement of OEF, CBF and $CMRO_2$. In addition, cerebrovascular reactivity (CVR) can also be extracted from the dc-fMRI data,[23,169] which itself is an important index of cerebrovascular health.[171,172] Figure 9 shows exemplar parametric maps produced by dc-fMRI.[23] The reproducibility of dc-fMRI measurements has been evaluated in two studies. Lajoie et al. reported a same-day CoV of 13.6% for OEF averaged across gray matter.[173] Merola et al. showed that the CoV of OEF averaged across gray matter was 6.7% and 10.5% for same-day and day-to-day test-retest experiments, respectively.[174] These CoV values are generally higher than some of the above-described methods. However, the advantage of the dc-fMRI method is that it provides spatially resolved maps of multiple hemodynamic and metabolic parameters in one scan. The sensitivity of dc-fMRI to OEF changes has been demonstrated in caffeine challenges.[117]

A key assumption in dc-fMRI is that hypercapnia and hyperoxia do not change $CMRO_2$. However, it is under debate whether these two challenges are truly "isometabolic".[47,48,58,93,113-116] Simulations have demonstrated that violation of the isometabolic assumption results in bias in OEF estimation.[175,176] Several approaches have been developed to account for possible alterations of $CMRO_2$ during gas challenges. Bulte et al. incorporated a fixed 10% reduction of $CMRO_2$ during hypercapnia in the model fitting.[23] Englund et al. proposed to measure global $Y_v$ together with CBF and BOLD to relax the isometabolic mandate, assuming that $Y_v$ changes induced by gas challenges are spatially uniform.[177] Driver et al. proposed to use graded hypercapnia to determine $M$ and dose-wise $CMRO_2$ alterations.[115]

A limitation of dc-fMRI is the need of gas challenges, which require complex set-ups and can lead to a considerable dropout rate for patients due to the discomfort of hypercapnia challenge.[164] A recent study has proposed to replace gas challenges with breath-hold modulations to quantify OEF using the BOLD signal model.[178,179]

Another limitation of the dc-fMRI method is that the reliability of the measure primarily hinges upon the quality of the ASL MRI data. ASL is known to suffer from low SNR even for basal perfusion measurement. The reliability of quantifying changes in CBF due to hypercapnia or hyperoxia requires further examination. Additionally, ASL signal may have confounding factors such as bolus arrival time, $T_2^*$, and labeling efficiency, which are also expected to alter during the physiological challenges.[121-124]

### 3.6. Comparison and combination among techniques

Table 1 shows a brief summary of the techniques described in previous sections and their strengths and weaknesses. In general, global OEF measurements have high SNR and some of them, such as TRUST and OxFlow, have been extensively tested and validated against gold standards.[20,41,44,45,49,94,101] For regional OEF measurement, a major challenge is the low SNR, because the local blood volume is very small and many techniques rely on complex model fittings that are sensitive to noise. Some have proposed to exploit machine learning methods to denoise the OEF maps and reduce the computational cost.[170,180,181] Future technical developments to improve the SNR is critical for the robustness of regional OEF measurement.

Several studies have compared the OEF values measured with different MRI-based techniques. For example, Barhoum et al. showed a significant correlation ($R^2$=0.50) between TRUST and OxFlow $Y_v$ measurements.[94] Significant correlations have also been reported between dc-fMRI and OxFlow[182] and between dc-fMRI and a QSM-based regional OEF method ($R^2$=0.39).[183] Overall, the correlations between different MRI methods were moderate and systematic differences in OEF quantification were observed.[94,111,183-185]

Combinations of different MRI OEF techniques have been proposed.[186,187] For example, Cho et al. combined the QSM model of the phase data and the qBOLD model of the magnitude data to estimate OEF from complex multi-echo GRE signals.[187-189] One advantage of this QSM+qBOLD method is the joint estimation of the unknowns including $Y_v$, $CBV_v$, and non-blood susceptibility,[187] thus eliminating the assumptions needed in previous QSM methods.[17,110-112,120] Figure 10 shows representative parametric maps generated by the QSM+qBOLD method.[187] Recently, this QSM+qBOLD method has been compared to $^{15}$O-PET.[128] Although the averaged OEF values across subjects were not significantly different between QSM+qBOLD and $^{15}$O-PET, there was no report of correlation between these two methods.[128]

## 4. Clinical applications of MRI-based OEF

In the following sections, we will highlight several clinical applications of the MRI-based OEF techniques, although they are not meant to be exhaustive.

### 4.1. OEF across the human lifespan

A number of MRI studies have investigated the evolution of OEF from fetuses to elderly individuals.

For fetuses, a few studies used QSM-based methods to measure $Y_v$ in the SSS in the fetal brain,[190-194] and showed that the median $Y_v$ across fetuses decreased from 67.5% in the second trimester to 60.8% in the third trimester.[193] Note that oxygen is delivered to the fetuses from the placenta through the umbilical vein, which was reported to have an average oxygenation of 84%.[195]

For newborn infants, several MRI techniques have been adapted to measure OEF in the neonatal brain, including $T_2$-based[57,62,196-198] and phase-based[199] methods. In healthy neonates, Liu et al. reported an average global OEF of 31.8±4.1%, which was similar to the adult OEF level.[197]

There is currently a paucity of human studies on age-related OEF changes in children of 1 to 18 years of age. A previous $^{15}$O-PET study found that OEF values in children were within the range of adult values regardless of the children's age.[200]

During adulthood, many MRI-based studies have reported an age-related increase in OEF,[2,49,74,201-203] which was accompanied by a reduction in CBF.[74,202,203] Results regarding $CMRO_2$ are mixed.[74,202-204] Increase in OEF with age has also been suggested in recent studies using non-MRI techniques, such as $^{15}$O-PET[205] and near-infrared spectroscopy[206-208] in human subjects.

**4.2. Cognitive impairment**

AD and vascular disease, as well as their co-occurrence, are the most common causes of cognitive impairment.[209] There are some suggestions that OEF is differentially affected by AD and vascular disease.[2] For example, AD pathology will lead to diminished neural activities, and thereby decreased glucose and oxygen metabolism.[210] Thus, a decreased OEF is expected in the presence of relatively intact blood supply. On the other hand, small-vessel vascular pathology will cause a reduction in blood supply[211,212] and result in an elevated oxygen extraction (Figure 11).[2] This notion is consistent with other recent MRI studies, which reported reduced OEF in cognitively impaired patients with minimal vascular risk factors.[3,164] In addition, it has been shown that among cognitively normal individuals, the carriers of the apolipoprotein-E4 gene, a major genetic risk factor for AD, manifested diminished global OEF.[75]

**4.3. Cerebral large and small vessel diseases**

In patients with unilateral stenosis/occlusion of major cerebral arteries, several MRI studies have reported an elevated OEF in the affected hemisphere,[126,213-217] consistent with the "misery perfusion" pattern observed in previous $^{15}$O-PET studies;[4,218-220] while a few other MRI studies found insignificant hemispheric difference in OEF.[221-225] MRI techniques have also been used to evaluate OEF in patients before or after carotid stenting/endarterectomy.[226,227]

In patients with Moyamoya disease, Watchmaker et al. found an elevated MRI-based global OEF,[228] while $^{15}$O-PET studies have reported mixed results.[229-231]

In patients with cerebral small vessel disease, a recent MRI study found an elevated OEF in white matter and watershed regions, while the OEF in gray matter was decreased.[232]

## 4.4. Stroke

In acute ischemic stroke patients, a few MRI studies have reported that OEF was higher in the affected hemisphere than the contralateral side.[233-235] A study by An et al. further attempted to delineate the stroke penumbra using MRI-based OEF and CBF measures.[236] Several other MRI studies have found a decreased $CMRO_2$[237-239] but increased OEF[240] in the ischemic core of acute stroke patients.

A few studies reported an elevated $R_2'$ in ischemic tissues in acute stroke,[241-244] but did not quantify OEF. Because $R_2'$ is proportional to the product of OEF and DBV (see Section 3.4), consideration of DBV is crucial to assess the extent of OEF changes using the $R_2'$ data.[241]

## 4.5. Sickle cell disease

A number of studies have investigated OEF changes in SCD but reported inconsistent results. Several groups have measured global OEF in SCD using $T_2$-based techniques,[5,54,81,228,245-250] but the OEF results relied on the calibration model used to convert blood $T_2$ to $Y_v$ (see Section 3.1.2).[251] Studies using the bovine calibration model[43] showed that OEF was higher in SCD patients than controls,[228,245] and OEF was reduced after blood transfusion.[246] In contrast, studies using the SCD-specific models[5,81] have either reported a reduced OEF in SCD patients,[5,247-249] or found no significant difference in OEF between SCD patients and controls.[54] One study used the neonatal calibration model,[78] and showed that elevated OEF was associated with impaired processing speed in SCD patients.[250]

Another group used ASE-based qBOLD methods and have consistently reported elevated whole-brain and regional OEF in children with SCD.[6,252-255] In addition, in children receiving chronic transfusion therapy, CBF and OEF were reduced after transfusion, as shown in Figure 12.[253]

Two studies used susceptibility-based techniques, one found reduced OEF in the SSS,[256] while another study showed elevated OEF in ICVs in SCD patients.[257]

The discrepancy in the literature suggests that the application of MRI-OEF techniques to pathological conditions with atypical RBC still presents some challenges. Future studies are needed to resolve the controversies in this field.

### 4.6. Brain injury

In neonates, hypoxic ischemic encephalopathy (HIE) is a leading cause of neonatal mortality and neurological disabilities.[258] A few MRI-based studies have reported that neonates with HIE had lower OEF than controls,[57,259] and neonates with severe HIE had even lower OEF than those with moderate HIE,[260] which is presumably due to lower oxygen consumption rate.

In children and adults, traumatic brain injury (TBI) is among the most severe types of injury in terms of fatality and long-term impairment.[261] Several MRI-based studies have suggested that OEF was reduced in patients with TBI.[262-266] In addition, in patients with TBI, a higher OEF predicted a better clinical outcome.[262]

### 4.7. OEF in physiological challenges

A number of studies have investigated the change of OEF and $CMRO_2$ under various physiological challenges, including hypercapnia,[47,58,93,113-115,267] hypocapnia (e.g., induced by hyperventilation),[110,113,268-270] hyperoxia,[48,116,271,272] hypoxia,[48,114,271,273-275] caffeine ingestion,[17,46,99,117,276,277] acetazolamide injection[99,248,278] and acute glucose ingestion.[279] Table 2 summarizes the OEF, CBF and $CMRO_2$ changes under different physiological challenges.

### 4.8. Other applications

Elevated OEF was reported in patients with end-stage renal disease,[280-282] hepatic encephalopathy,[283] systemic lupus erythematosus,[284] refractory epilepsy[285] and chronic cannabis usage.[286] Increased OEF was also found in children with primary nocturnal enuresis[287] and in preterm neonates with anemia.[288]

Reduced OEF was shown in patients with MELAS syndrome[289] and multiple sclerosis,[290,291] as well as in neonates with punctate white matter lesions.[292]

OEF and $CMRO_2$ have also been studied in patients with glioma,[7,293-299] mountain sickness,[274] cocaine addiction,[300] anorexia,[301] bipolar disorder,[302] metabolic disorder[303] and obstructive sleep apnea.[304,305]

Besides disease-related changes, alterations of OEF and $CMRO_2$ have also been observed during natural sleep[306] and after fatiguing aerobic exercise.[307]

## Conclusions

This work provides an overview of emerging MRI techniques for OEF measurement. We reviewed the methodological aspects of different categories of MRI techniques, and discussed their advantages and limitations. We also briefly reviewed the applications of MRI-based OEF measurements in various physiological or pathological conditions. In summary, MRI-based OEF techniques provide a non-invasive assessment of the brain’s function and metabolism, and have strong potentials in a variety of basic science and clinical applications.

Table 1. Summary of MRI OEF techniques.

| Method | Exemplar pulse sequences | Pros | Cons |
|---|---|---|---|
| $T_2$-based global | • TRUST[40]<br>• $T_2$-TRIR[57]<br>• High-resolution $T_2$ mapping[55] | • High SNR<br>• Short scan time<br>• Straightforward data processing<br>• Excellent reproducibility | • Lack of spatial specificity<br>• Dependence on $T_2$-Y calibration model |
| $T_2$-based regional | • TRUPC[62]<br>• QUIXOTIC[65]<br>• VSEAN[66] | • Vessel-specific or voxel-wise mapping of OEF | • Low SNR<br>• Dependence on $T_2$-Y calibration model |
| Phase-based global | • OxFlow[20]<br>• Flow-compensated multi-echo GRE[19] | • High SNR<br>• Very short scan time<br>• Excellent reproducibility | • Lack of spatial specificity<br>• Restriction on vessel orientation and shape<br>• Dependent on manual drawings of vessel and tissue ROI<br>• $\Delta\chi_{do}$ of special RBC not fully characterized |
| QSM-based regional | • Flow-compensated multi-echo GRE[15-17] | • Vessel-specific or voxel-wise mapping of OEF | • Susceptible to partial volume effect<br>• Need of prior assumptions<br>• Relatively long scan time |
| qBOLD | • GESSE[133]<br>• ASE[134] | • Voxel-wise mapping of OEF | • Complex signal model<br>• Confounding factors such as macroscopic field inhomogeneity<br>• Relatively long scan time |
| dc-fMRI | • Dual-echo ASL[22,24] | • Voxel-wise mapping of OEF, CBF, $CMRO_2$ and CVR | • Need of gas challenges<br>• Complex signal model<br>• Low SNR due to the need to measure ASL during baseline and challenged states<br>• Long scan time |

Table 2. OEF, CBF and $CMRO_2$ under physiological challenges.

| Physiological challenge | CBF | OEF (or $AVDO_2$) | $CMRO_2$ |
|---|---|---|---|
| Hypercapnia | Increased[47,58,93,113,114,267] | Decreased[47,58,93,113,114,267] | Mixed literature reporting decreased[47,114,115,267] or unchanged[93,113] |
| Hypocapnia | Decreased[110,113,268-270] | Increased[110,113,268-270] | Unchanged[110,113,268-270] |
| Hyperoxia | Decreased[48,308] or unchanged[116,271,272] | Unchanged[48,116,271,272] | Mixed literature reporting decreased,[48] or unchanged[116,271,272] |
| Hypoxia | Increased[48,114,271,273,274] or unchanged[275] | Unchanged[48,271,273-275] | Mixed literature reporting increased[48,114,273,274] or unchanged[271,275] |
| Caffeine | Decreased[17,46,99,117] | Increased[17,46,99,117] | Mixed literature reporting decreased,[117] increased[276] or unchanged[46,277] |
| Acetazolamide | Increased[99,248,278] | Decreased[99,248,278] | Unchanged[99,248,278] |
| Acute glucose ingestion | Unchanged[279] | Decreased[279] | Decreased[279] |

**Figures**

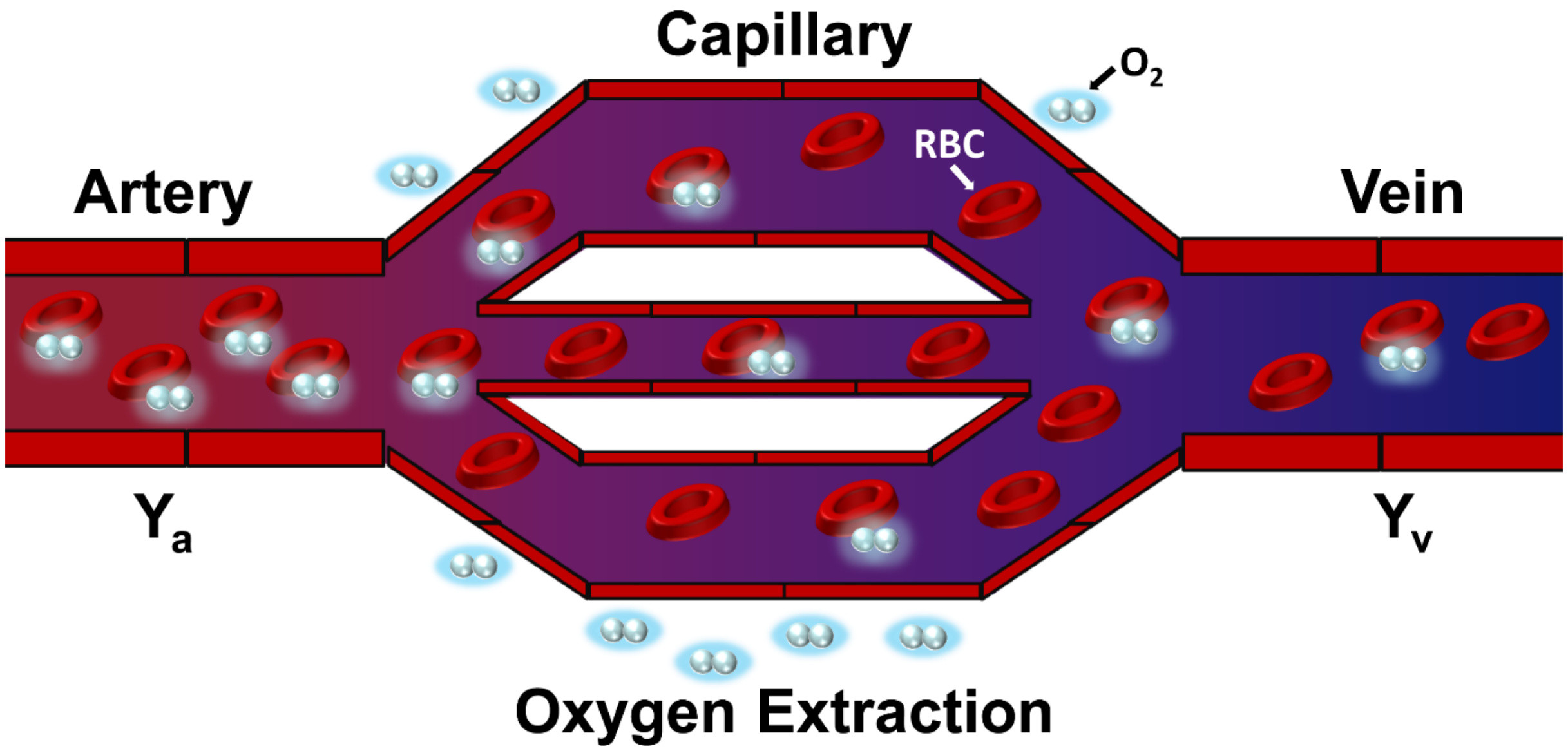

Figure 1. Illustration of oxygen extraction at the capillary bed.

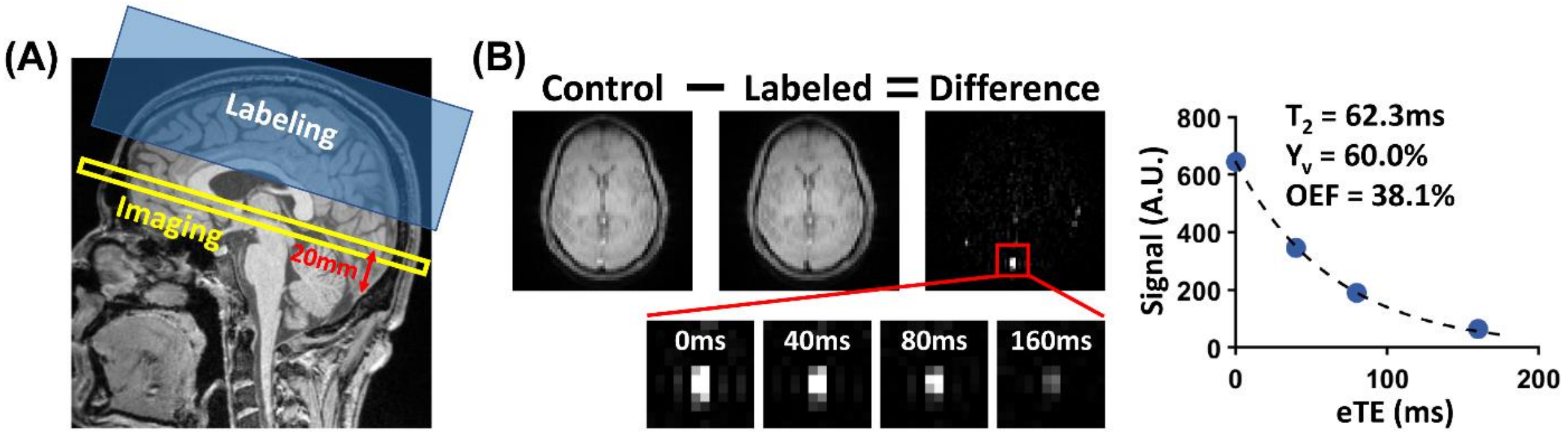

Figure 2. Illustration of TRUST MRI. (A) Position of TRUST MRI. Imaging slice (yellow box) is placed to be approximately parallel to the anterior-commissure–posterior-commissure line and 20mm above sinus confluence. Blue box represents the labeling slab. (B) Representative TRUST data. Subtraction between control and labeled images yields strong venous blood signal in the SSS (red box). The scatter plot shows venous signal as a function of eTEs. The resulting blood $T_2$, $Y_v$, and OEF are also shown.

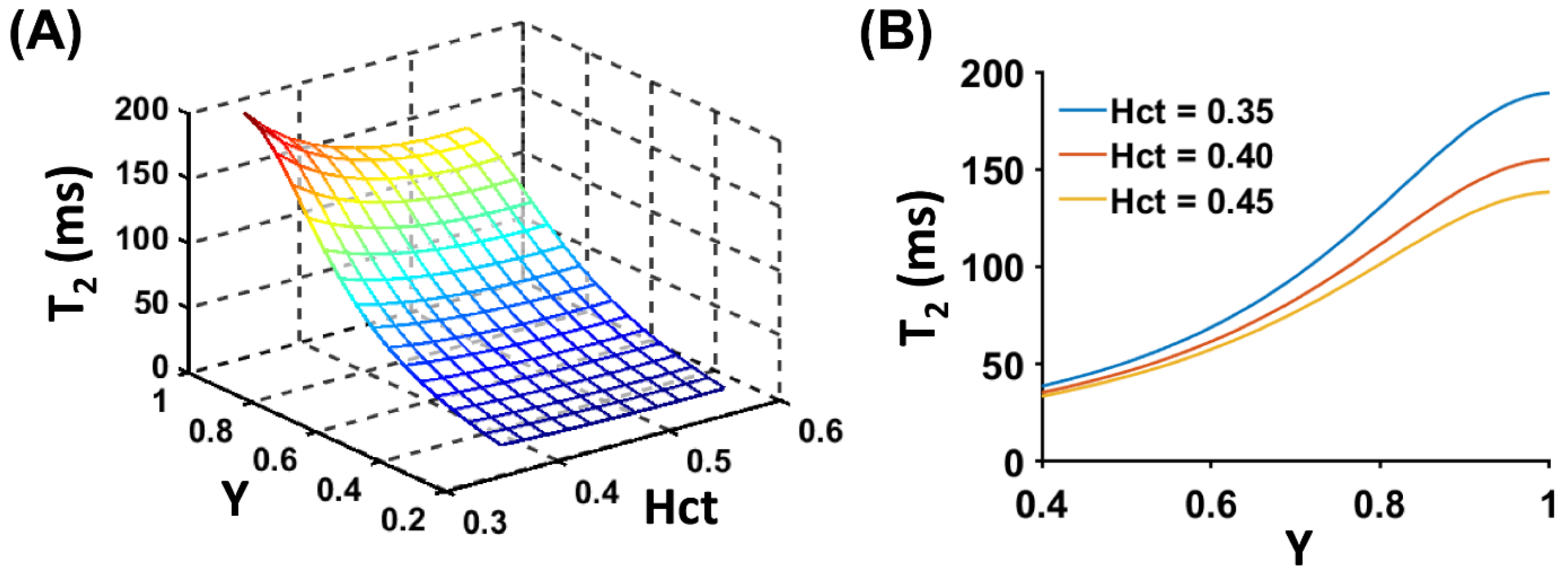

Figure 3. The bovine $T_2$-Y calibration model. (A) 3D mesh plot showing the dependence of blood $T_2$ on Y and Hct; (B) Exemplar $T_2$-Y conversion curves extracted from the 3D mesh plot. With a fixed Hct, $T_2$ increases monotonically with Y; with a fixed Y, a lower Hct leads to a higher $T_2$.

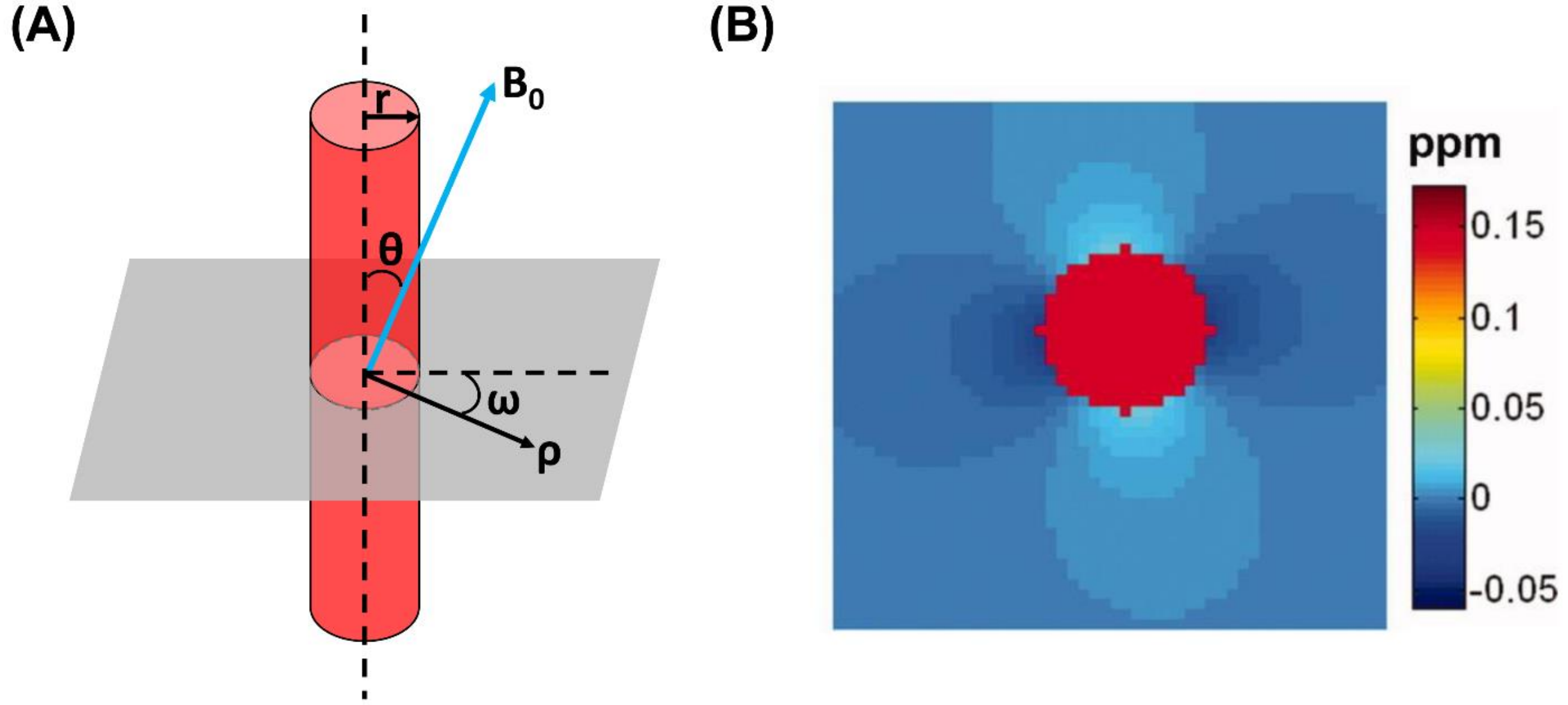

Figure 4. Blood-induced field of a cylindric vessel. (A) Illustration of the spatial relationship between the vessel and the variables in Eqs. [8,9]; (B) Simulated blood-induced field in a plane perpendicular to the vessel (gray parallelogram in [A]), assuming $Y_v$=65% and θ=20.1°. The induced field is represented in parts per million (ppm) with respect to $B_0$. It is homogenous inside the vessel (red circle) and decays rapidly with the distance from the vessel. Adapted from Li et al.[90] with permission.

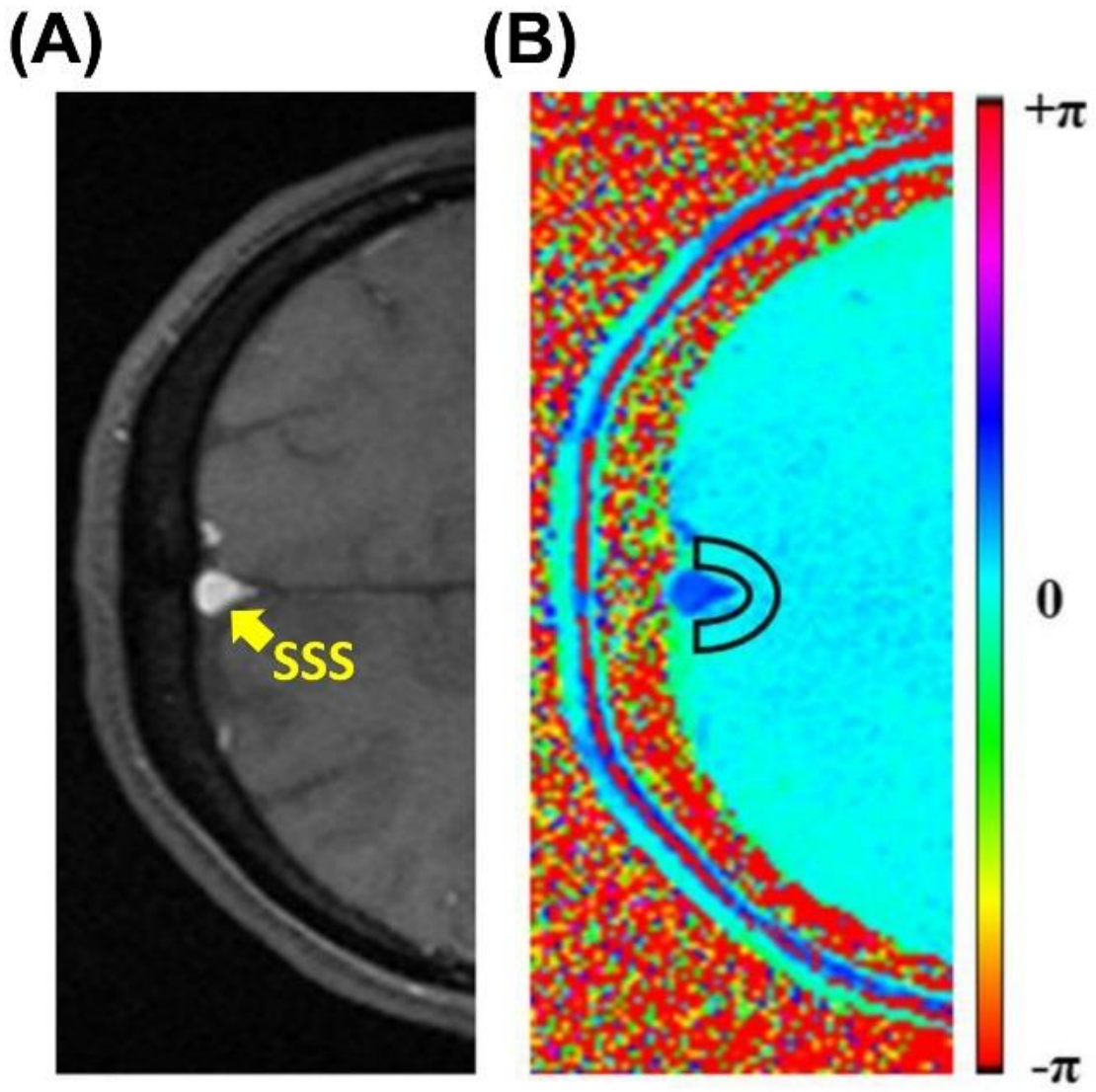

Figure 5. Exemplar data of phase-based OEF. Magnitude (A) and phase (B) images at the level of SSS (yellow arrow). The region-of-interest for extravascular tissue is denoted by the black box. Adapted from Barhoum et al.[94] with permission.

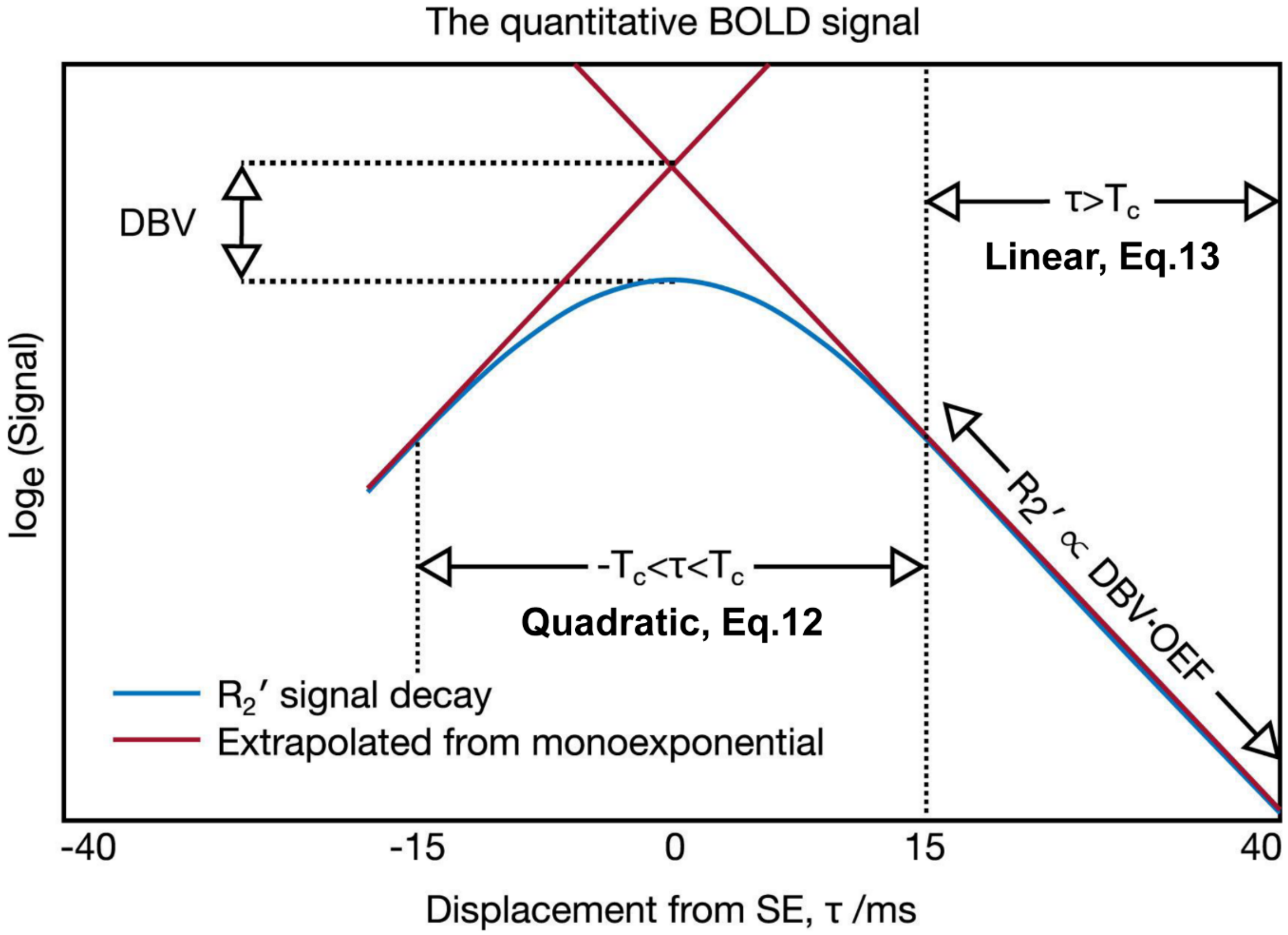

Figure 6. Schematic of the qBOLD model describing the transverse MR signal decay in the presence of a blood vessel network. $R_2'$ is inferred from the monoexponential regime ($T_c = 1.5/\delta\omega$, Eq. [13]) and DBV is inferred from the mismatch between the linear intercept of this fit and the spin echo signal ($\tau$=0ms). Baseline OEF can then be estimated by combining these two measurements. Adapted from Stone et al.[130] with permission.

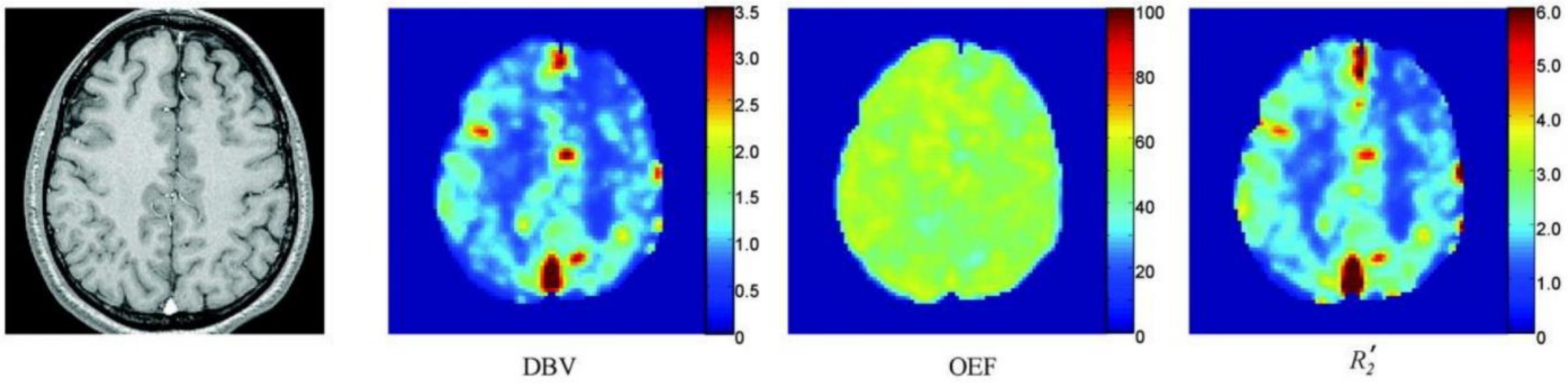

Figure 7. qBOLD parametric maps from a representative subject. DBV, OEF and $R_2'$ maps are presented alongside an anatomic image. Adapted from He et al.[133] with permission.

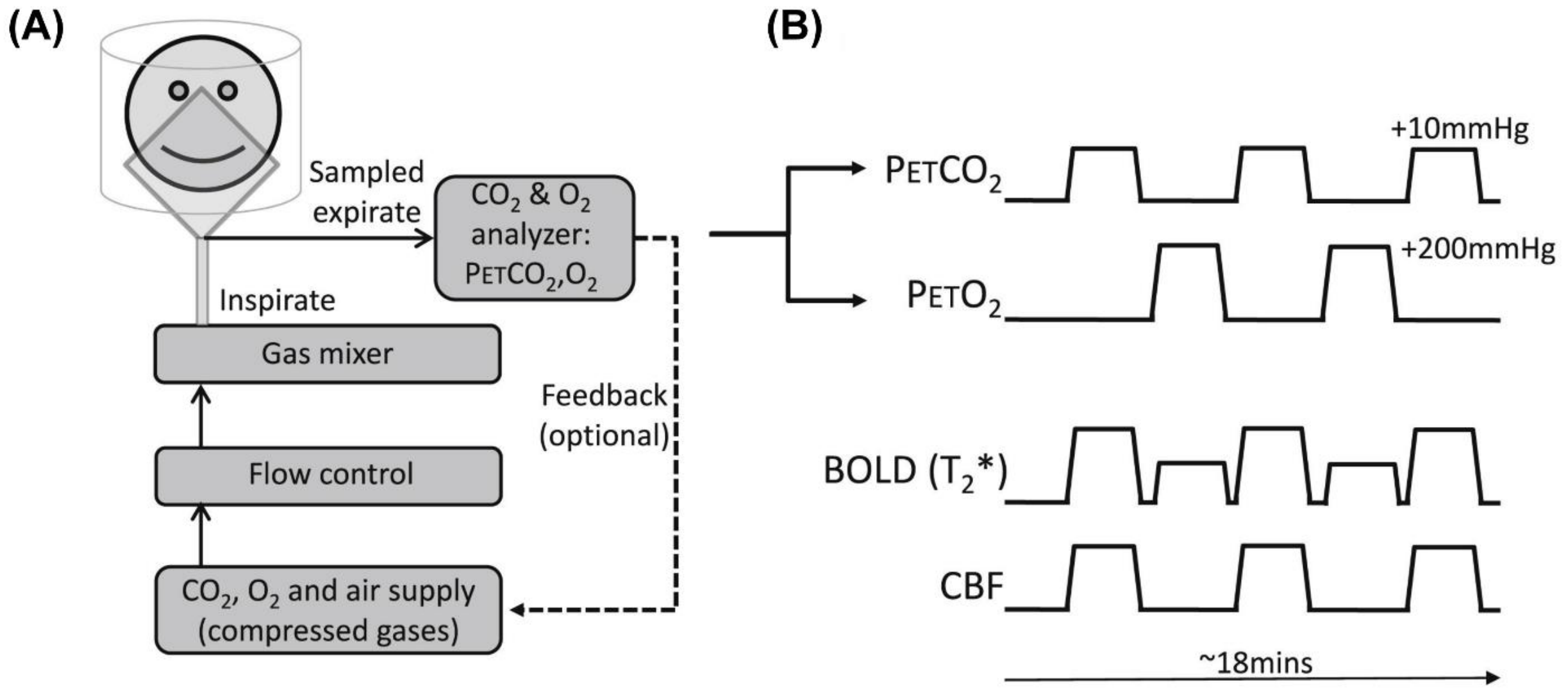

Figure 8. Schematic of dc-fMRI experiment. (A) Respiratory manipulations. (B) The corresponding idealized partial pressure of end-tidal $CO_2$ ($P_{ET}CO_2$), $P_{ET}O_2$, BOLD and CBF time-courses. Reproduced from Germuska et al.[163] with permission.

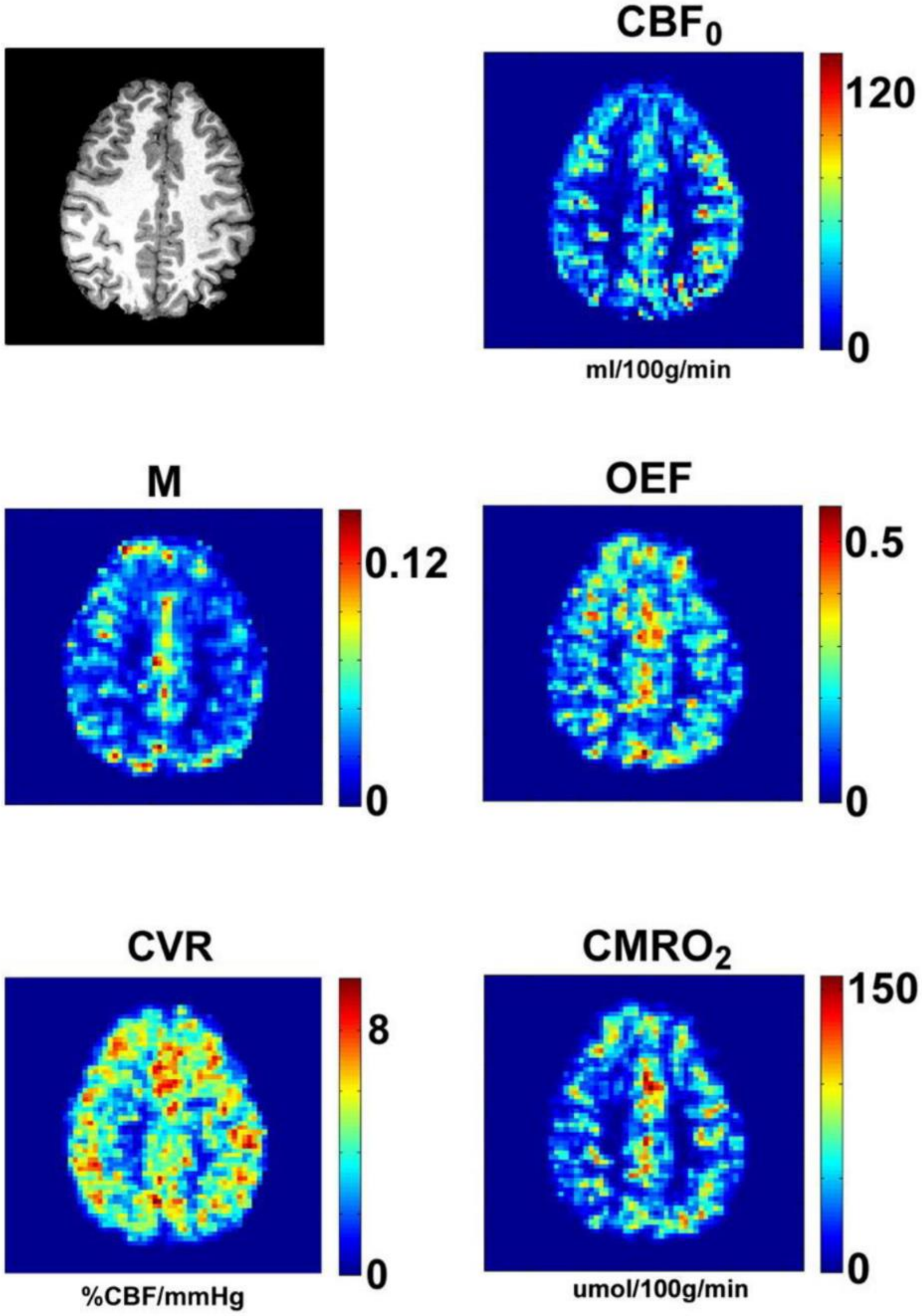

Figure 9. Exemplar parametric maps obtained with dc-fMRI. Adapted from Bulte et al.[23] with permission.

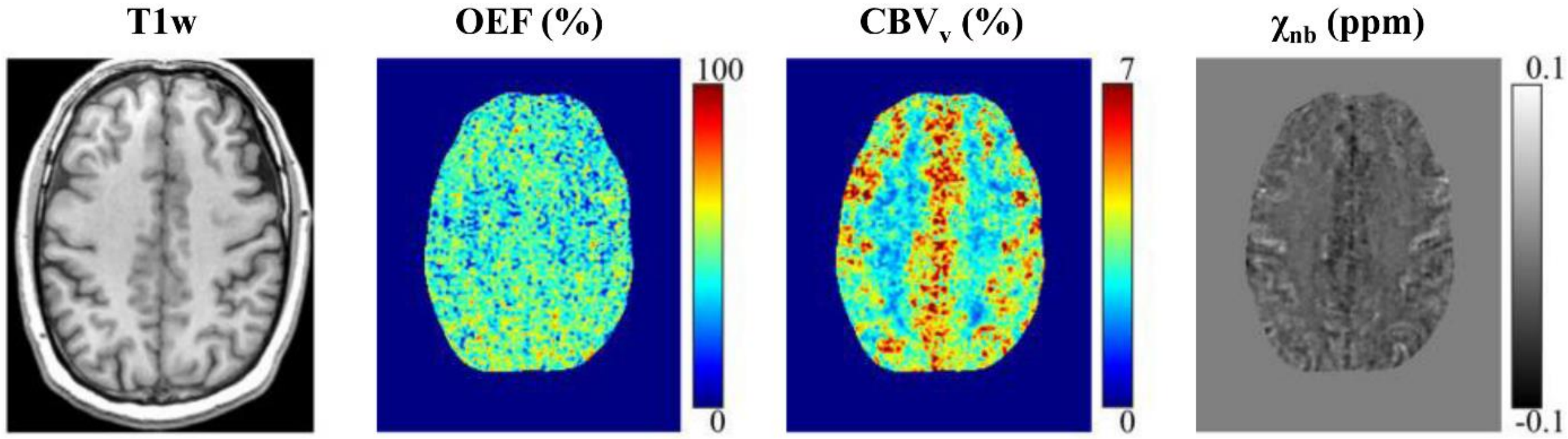

Figure 10. Representative OEF, venous cerebral blood volume ($CBV_v$) and non-blood susceptibility ($\chi_{nb}$) maps generated by the QSM+qBOLD model. Adapted from Cho et al.[187] with permission.

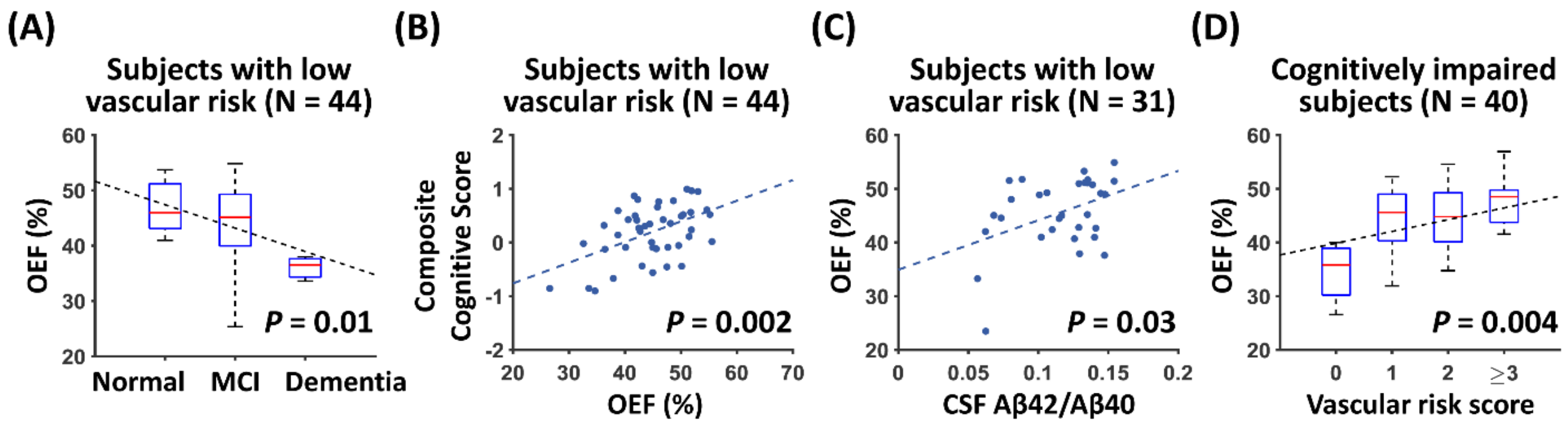

Figure 11. OEF values in elderly individuals, stratified by vascular risk and cognitive diagnosis. In individuals with low vascular risk: (A) patients with mild cognitive impairment (MCI) or dementia had lower OEF; (B) a lower OEF was associated with a worse cognitive performance; and (C) a lower OEF was associated with a lower CSF amyloid-42/amyloid-40 (Aβ42/Aβ40) ratio (corresponding to a greater amyloid burden in the brain). (D) In cognitively impaired patients (MCI or dementia), a higher OEF was associated with a greater vascular risk. Adapted from Jiang et al.[2] with permission.

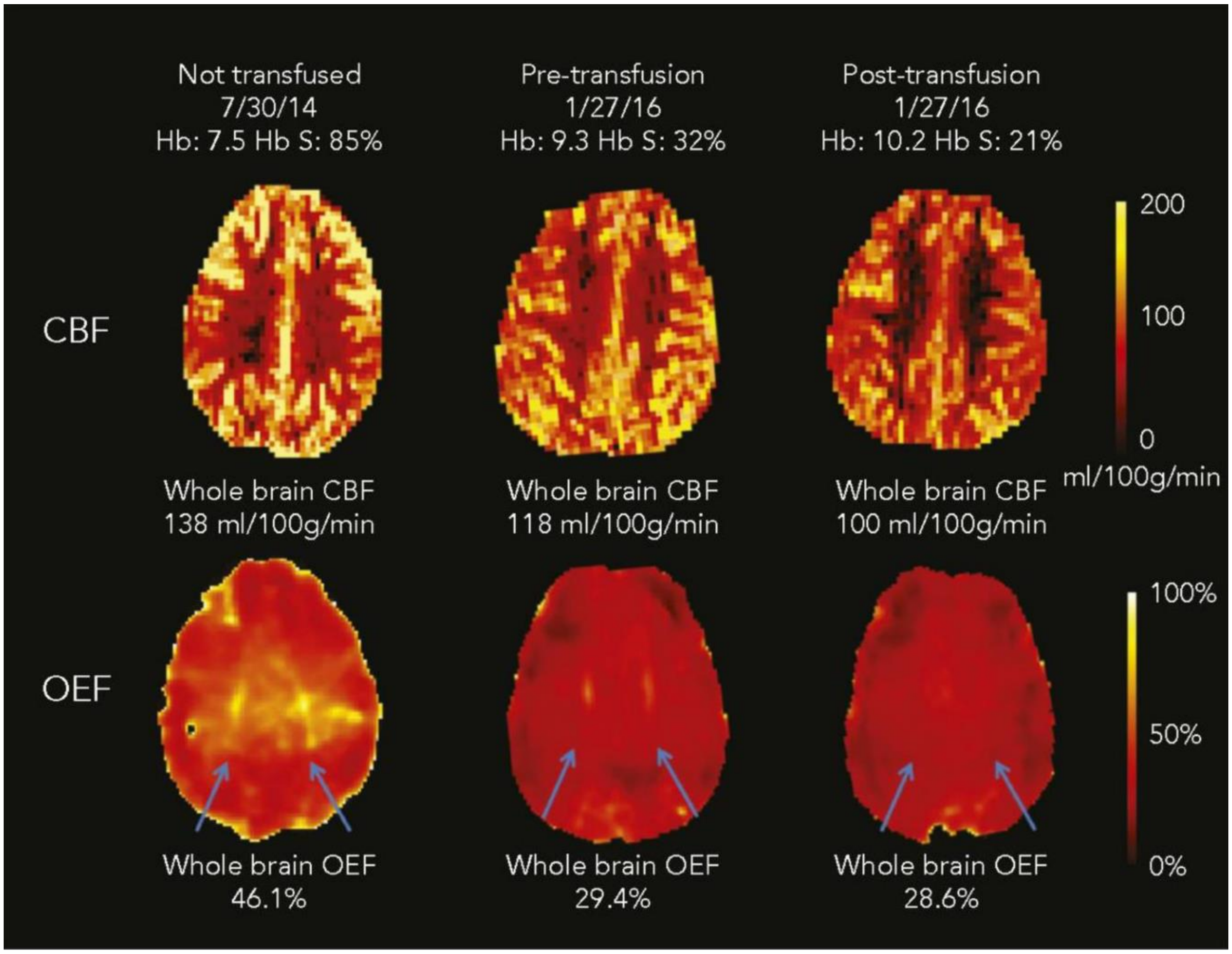

Figure 12. CBF and OEF maps from a child with SCD. This 7-year-old boy underwent an MRI scan before chronic transfusion therapy (CTT) initiation and another two MRI scans before and after an exchange transfusion. The whole-brain CBF was highest at his first scan; after 17 months of CTT, his pretransfusion CBF was lower than the initial scan and the posttransfusion CBF was further decreased. OEF was highest at the first scan, but was dramatically reduced pretransfusion and further decreased posttransfusion. Reproduced from Guilliams et al.[253] with permission.